# Low-loss thin-film periodically poled lithium niobate waveguides fabricated by femtosecond laser photolithography

GUANGHUI ZHAO,[1,4] JINTIAN LIN,[1,5,6,12] RENHONG GAO,[2,3,11] QIFENG HOU,[1,6] JIANGLIN GUAN,[2,3] CHUNTAO LI,[2,3] XINZHI ZHENG,[2,3] MINGHUI LI,[1,5] XIAOCHAO LUO,[1,5] YINGNUO QIU,[1,5] LINGLING QIAO,[1] MIN WANG,[2] AND YA CHENG[1,2,3,4,7,8,9,10,13]

[1]*State Key Laboratory of Ultra-intense laser Science and Technology, Shanghai Institute of Optics and Fine Mechanics, Chinese Academy of Sciences, Shanghai 201800, China*
[2]*The Extreme Optoelectromechanics Laboratory (XXL), School of Physics and Electronic Science, East China Normal University, Shanghai 200241, China*
[3]*State Key Laboratory of Precision Spectroscopy, East China Normal University, Shanghai 200062, China*
[4]*School of Physical Science and Technology, ShanghaiTech University, Shanghai 200031, China*
[5]*Center of Materials Science and Optoelectronics Engineering, University of Chinese Academy of Sciences, Beijing 100049, China*
[6]*School of Physical Sciences, University of Science and Technology of China, Hefei 230026, China*
[7]*Shanghai Research Center for Quantum Sciences, Shanghai 201315, China*
[8]*Hefei National Laboratory, Hefei 230088, China*
[9]*Collaborative Innovation Center of Extreme Optics, Shanxi University, Taiyuan 030006, China*
[10]*Collaborative Innovation Center of Light Manipulations and Applications, Shandong Normal University, Jinan 250358, China*
[11]*rhgao@phy.ecnu.edu.cn*
[12]*jintianlin@siom.ac.cn*
[13]*ya.cheng@siom.ac.cn*

**Periodically poled lithium niobate on insulator (PPLNOI) ridge waveguides are critical photonic components for both classical and quantum information processing. However, dry etching of PPLNOI waveguides often generates rough sidewalls and variations in the etching rates of oppositely poled lithium niobate ferroelectric domains, leading a relatively high propagation losses (0.25–1 dB/cm), which significantly limits net conversion efficiency and hinders scalable photonic integration. In this work, a low-loss PPLNOI ridge waveguide with a length of 7 mm was fabricated using ultra-smooth sidewalls through photolithography-assisted chemo-mechanical etching (PLACE) followed by high-voltage pulse poling with low cost. The average surface roughness was measured at just 0.27 nm, resulting in record-low propagation loss of 0.106 dB/cm in PPLNOI waveguides. Highly efficient second-harmonic generation was demonstrated with a normalized efficiency of 1643% $W^{-1}\cdot cm^{-2}$ without temperature tuning, corresponding to a conversion efficiency of 805%/W, which is closed to the best conversion efficiency (i.e., 814%/W) reported in nanophotonic PPLNOI waveguide fabricated by expensive electron-beam lithography followed by dry etching. The absolute conversion efficiency reached 15.8% at a pump level of 21.6 mW. And the normalized efficiency can be even improved to 1742% $W^{-1}\cdot cm^{-2}$ at optimal temperature of 59º.**

In recent years, the commercialization of thin-film lithium niobate on insulator (LNOI) wafers and advancements in ultra-low-loss waveguide fabrication technologies [1,2] have established LNOI as a highly promising material platform for multi-functional integrated photonics [3-5]. This is attributed to the strong optical confinement combined with the superior optical properties of LNOI, including low material loss, broad transparency window, strong electro-optic effect, and significant second-order nonlinear optical responses. Leveraging these advantages, various photonic devices such as high-speed electro-optic modulators, efficient nonlinear frequency converters, and entangled photon pair sources have been demonstrated on LNOI platforms [1-5,6-11]. Among these applications, nonlinear integrated photonic devices based on ferroelectric domain engineering have garnered significant interest for both classical and quantum information processing [12-23]. As a typical ferroelectric material, lithium niobate (LN) allows post-fabrication domain inversion through poling processes, enabling quasi-phase matching (QPM) to significantly enhance nonlinear conversion efficiency via periodic poling.

The fabrication of PPLNOI waveguides involves two essential processes: ferroelectric domain inversion and waveguide etching. These can be performed in sequence [13-16,18-24] or inversely [17,25,26]. However, periodically poled lithium niobate (PPLN) exhibits selective anisotropic etching issues during reactive ion etching and other processes like RCA cleaning [25]. The accessible nonlinear conversion efficiency is highly susceptible to propagation loss, necessitating low propagation loss for high conversion efficiency. Conventional fabrication procedures involving periodic domain inversion followed by waveguide etching result in significant propagation losses (~1 dB/cm) in the telecom band, severely limiting conversion efficiency by extending the waveguide length and photonic integration scalability.

To address this challenge, an alternative approach involves performing periodic poling after waveguide etching. This method only introduces a slight additional propagation loss (from 0.23 dB/cm to 0.25 dB/cm in the telecom band before and after periodic poling [25]), generating the lowest loss PPLNOI waveguides [25]. The primary cause of this propagation loss is the relatively rough sidewall of dry-etched waveguides with an average surface roughness of ~1 nm, where expensive electron-beam lithography should be employed. To further improve net conversion efficiency and enable large-scale scalable photonic integration, it is essential to fabricate PPLNOI waveguides with ultra-smooth sidewalls to reduce propagation losses.

In this work, a low propagation-loss PPLNOI ridge waveguide (7 mm long) was fabricated via femtosecond laser photolithography-assisted chemo-mechanical etching (PLACE) [27] followed by periodic poling. This method avoids sidewall morphology degradation and periodic parameter deviations caused by anisotropic etching in conventional methods. The resulting PPLNOI ridge waveguide features ultra-smooth surfaces and sidewalls with an average surface roughness of 0.27 nm, significantly reducing scattering loss. The propagation loss reaches 0.11 dB/cm, representing a two-fold improvement in PPLNOI waveguides [25]. Highly efficient second harmonic generation (SHG) was achieved, with a normalized conversion efficiency of 1643%·W$^{-1}$·cm$^{-2}$, and a slope (i.e., conversion efficiency) of 805%/W (without deducting propagation loss). The on-chip conversion efficiency is closed to the record (i.e., 814%/W) reported in nanophotonic single-periodic PPLNOI waveguides fabricated by expensive electron-beam lithography followed by dry etching [28]. This technique offers a reliable pathway for large-scale, high-density nonlinear photonic circuits at low cost.

A commercially available X-cut LNOI wafer (Shanghai Novel Si Integration Technology Co.) was used in this experiment, which consists of a 500-nm-thick LN thin-film layer, a 4.7-μm-thick silicon dioxide (SiO$_2$) buffer layer and an LN handle. The PPLNOI ridge waveguide was fabricated by waveguide etching using PLCAE technique followed by periodic poling, as schematically illustrated in Figure 1. First, a 200-nm-thick chromium (Cr) layer was deposited on the thin-film lithium niobate (TFLN) via magnetron sputtering. Second, a femtosecond pulsed laser was used to selectively ablate the Cr layer for generating strip patterns with a width of 1 μm, oriented to Y axis of the crystal. Third, chemo-mechanical polishing (CMP) was employed to etch the exposed TFLN with a depth of 210 nm, resulting in pattern transferring from Cr mask pattern to the TFLN. The residual Cr mask was removed via wet etching after the fabrication of LNOI ridge waveguide. Fourth, a multilayer stack of 100 nm Cr, 200 nm Au, and 100 nm Ti was sequentially sputtered onto the pre-fabricated ridge waveguide sample. Fifth, femtosecond laser direct writing was used to pattern the Ti layer by ablation, followed by wet etching the unprotected Au layer with aqua regia and removing the Ti mask and Cr, so as to produce comb-shaped microelectrodes. Sixth, to achieve domain inversion, several high-voltage pulses (400 V, 5 ms duration) were applied to the comb-shaped electrodes, generating periodic poling in the ridge waveguide region. The spacing between the electrode pair was approximately 7 μm, yielding an electric field strength of ~51.7 kV/mm, which exceeds the coercive field of bulk LN crystals (21 kV/mm), ensuring effective domain inversion. Finally, the residual Cr-Au electrodes were subsequently removed via chemical etching, resulting in the fabrication of the PPLNOI ridge waveguides.

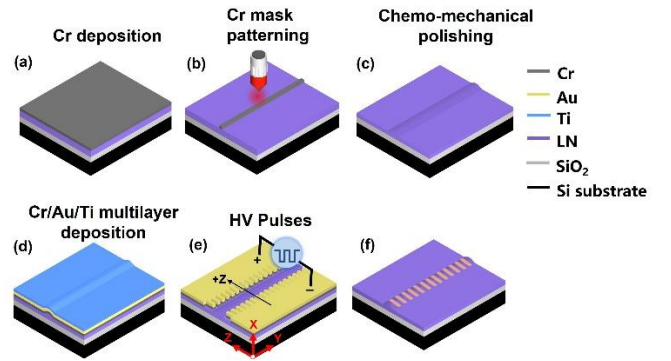

**Fig. 1.** Schematic of the fabrication the PPLNOI waveguide.

The optical microscope image of the fabricated PPLNOI ridge waveguide shows an ultra-smooth surface, as depicted in Fig. 2(a). And the scanning electron microscope (SEM) image of the cross section of ridge waveguide is shown in Fig. 2(b). The average surface roughness was measured to be only 0.27 nm by atomic force microscopy, as shown in the inset of Fig. 2(e). Consequently, the propagation loss was measured to be 0.106 dB/cm in the telecom band, by measuring the intrinsic Q factor of a racetrack PPLNOI microring with a physical length of 2 mm, which was $3.1 \times 10^6$. The propagation loss is the state of the art reported in PPLNOI waveguides, thanks to the ultra-smooth sidewall [27] and the elimination of anisotropic etching of periodic poling microstructures [25].

Moreover, this ridge waveguide only supports single-mode guiding in the telecom band, which is very important for high efficiency SHG by maximizing the overlap integral between the fundamental and the second harmonic modes. Since the poling period $\Lambda$ should provide optical momentum for quasi-phase matched SHG, it is necessary to calculate the effective refractive indices of the fundamental and the second harmonic modes. And the poling period was determined by the formula $\Lambda = \lambda_{2\omega}/(n_{2\omega} - n_\omega)$. Here, $\lambda_{2\omega}$, $n_{2\omega}$, and $n_\omega$ are the wavelength of the second harmonic in vacuum, the refractive indices of the second harmonic and the pump modes, respectively. The transverse-electrically (TE) polarized modal profiles are simulated by the finite element analysis method, as depicted in Fig. 2(c). And the effective refractive indices of both the 1550-nm pump and its second-harmonic modes are calculated to 1.8656 and 2.0797, respectively. Figure 2(d) illustrates the calculated QPM period as a function of the pump wavelength. As the pump wavelength redshifts, the required poling period gradually increases. When the pump wavelength was set at 1550 nm, the corresponding poling period should be 3.62 μm. After the fabrication of the PPLNOI ridge waveguide, the ferroelectric domain inversion was non-destructively characterized using confocal second harmonic (SH) microscopy, as depicted in Fig. 2(e), where $P_s$ labeled with a white line represents the spontaneous polarization orientation of LN. And the black line represents the domain inversion.

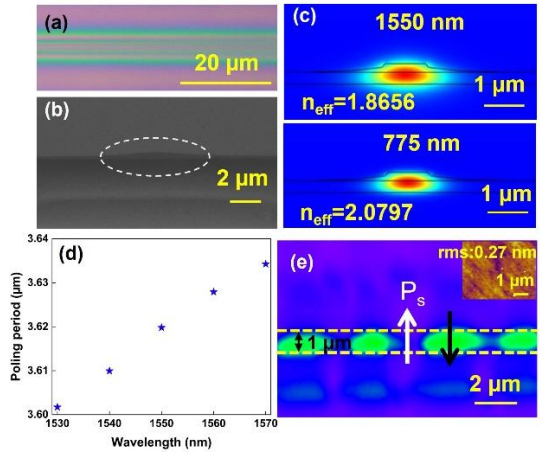

**Fig. 2.** Optical micrograph of the single-mode PPLNOI waveguide. (b) The SEM image of the cross section of the ridge waveguide. (c) Mode field profiles and effective refractive indices of the fundamental and second harmonic modes in the waveguide. (d) Simulation of optimum poling period varied with pump wavelengths. (e) Image of the domain inversion structure of the waveguide recorded using the confocal SH microscopy, where the waveguide was labeled with yellow lines with dashes. Inset: atomic force microscope (AFM) image of the waveguide.

To achieve SHG in the PPLNOI waveguide, a telecom-band tunable continuous-wave laser (Model: TLB-6728, New Focus Inc.) amplified with an erbium-doped fiber amplifier (EDFA) was employed as the pump source. The schematic of the experimental setup is shown in Figure 3. An inline fiber polarization controller was exploited to ensure the pump light was TE-polarized during the experiment. A lensed fiber was used for end-face coupling to inject pump light into the ridge waveguide. And an optical microscope imaging system which consists of an objective lens with numerical aperture of 0.25 and a charge coupled device (CCD) camera was mounted above the device under text (i.e., the waveguide chip, denoted as DUT) to monitor the coupling. Successively, the generated second harmonic signal was coupled out of the waveguide using another lensed fiber and sent to an optical spectrum analyzer (denoted as OSA, Model: YOKOGAWA AQ6370D) to monitor the second-harmonic signal and record its power. And the waveguide chip was placed on a thermoelectric heater for tuning the temperature. Here, the fiber-to-chip coupling losses of ~11 dB/facet and ~24 dB/facet have been calibrated and extracted by measuring the linear optical transmission at both pump and second harmonic wavelengths, respectively.

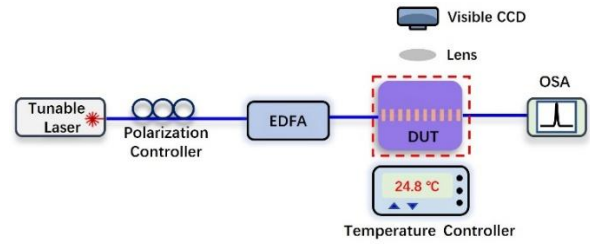

**Fig. 3.** Experimental setup for testing the PPLNOI waveguide.

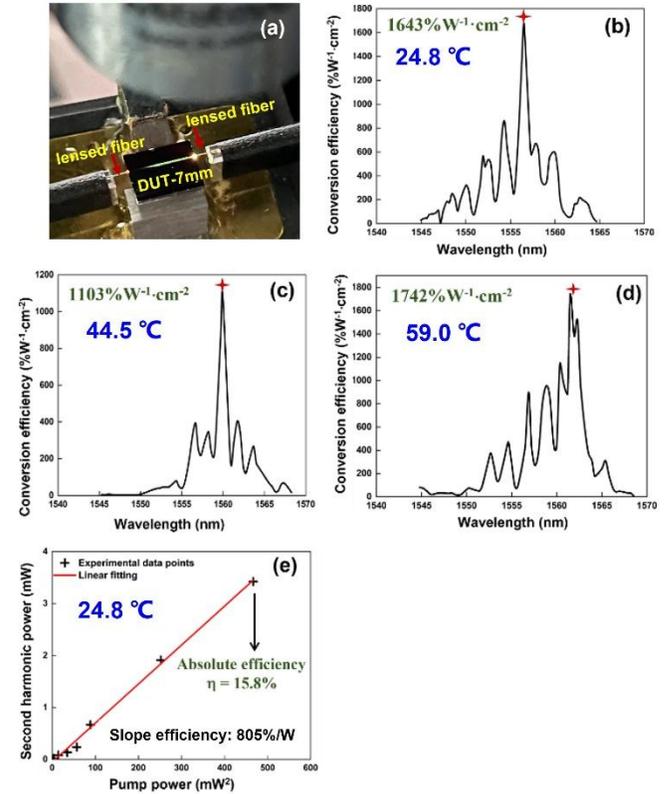

**Fig. 4.** Bright SHG generated in the waveguide captured by a smartphone. Normalized conversion efficiency of second harmonic as a function of pump wavelength at (b) 24.8°C, (c) 44.5°C, and (d) 59.0°C. (e) Quadratic power dependence of the second harmonic signal on the pump light at 24.8°C.

The pump wavelength was scanned from 1545 nm to 1565 nm, and the generated second harmonic signal was record by the OSA. When the temperature was set to room temperature (24.8°C), the normalized SHG conversion efficiency dependence on pump wavelength is plotted in Fig. 4(b). A highest normalized SHG conversion efficiency was measured as 1643 %W$^{-1}$·cm$^{-2}$ at 1556.56 nm pump. And bright SHG is visible to the naked eye, which was captured with a camera of a smart mobile phone, as shown in Fig. 4(a). Furthermore, the output power of second harmonic varied with pump powers was measured. Figure 4(e) illustrates a linear relationship between the SHG output power and the square of the pump light, where a slope of 805%/W was extracted. The absolute SHG conversion efficiency reaches a maximum of 15.8% (calculated according to the formula $\eta_{SHG} = P_{SHG}/P_{FW}$ where $P_{SHG}$ and $P_{FW}$ represent second harmonic and pump powers, respectively) at a pump level of 21.6 mW, corresponding to 3.4 mW SHG output.

Moreover, to improve the conversion efficiency, it is necessary to perform temperature tuning of the PPLNOI waveguide to improve phase match using the temperature heater. Typical experimental data was shown in Figs. 4(c) and 4(d), which were measured at 44.5°C and 59°C, respectively. The optimal phase-matching wavelength exhibited a red-shift with increasing temperature, corresponding to 1559.85 nm at 44.5°C and 1561.51 nm at 59°C. And the normalized conversion efficiencies were measured to 1103%W$^{-1}$cm$^{-2}$, and 1742%W$^{-1}$cm$^{-2}$, respectively.

In this work, a 7-mm-long low-loss PPLNOI waveguides has been fabricated by waveguide etching and microelectrode poling using the PLCAE technique to achieve efficient quasi-phase-matched SHG. Thanks to the ultra-smooth sidewalls of the waveguides produced by chemo-mechanical polishing and the elimination of anisotropic etching of PPLNOI waveguides in conventional fabrication processes, high normalized conversion efficiency has been demonstrated as 1643·W$^{-1}$·cm$^{-2}$ with a slope of 805%W$^{-1}$. Since the PLACE technique allows large-scale photonic integration with low propagation loss [29], our work opens up new opportunities for future quantum and classical photonic applications at low cost.

**Acknowledgment.** We thank Miss Lei Yang of East China Normal University for helping us to acquire the image of the domain inversion structure of the waveguide.